\input{aipcheck}

\documentclass[
    ,final            
  ,numberedheadings 
  ]
  {aipproc}

\layoutstyle{8x11single}

\begin{document}

\title[Geometry of Entanglement Sudden Death]{Geometry of Entanglement Sudden Death: Explicit Examples}


\author{Raphael C. Drumond}{
  address={Departamento de F\'isica, ICEx, UFMG\\
  Caixa Postal 702, Campus Pampulha, Belo Horizonte, MG, Brazil}
  email={raphaeld@fisica.ufmg.br}
}

\author{Marcelo O. Terra Cunha}{
  address={Departamento de Matem\'atica, ICEx, UFMG\\
  Caixa Postal 702, Campus Pampulha, Belo Horizonte, MG, Brazil}
  email={tcunha@mat.ufmg.br}
}


\begin{abstract}
 In open quantum systems, entanglement can vanish faster than coherence. This phenomenon is usually called sudden death of entanglement. In
[M.~O. Terra Cunha, {\it{New J. Phys.}} {\bf{9}}, 237 (2007)] a
geometrical explanation was offered and a classification of all
possible scenarios was given. Some classes were exemplified, but it
was still an open question whether there were examples for the other
ones. This was solved in [R.~C. Drumond and M.~O. Terra Cunha,
arXiv:0809.4445v1]. Here we briefly review the problem, state our
results in a precise way, discuss the generality of the approach,
and add some speculative desirable generalizations.
\end{abstract}

\maketitle


\section{Introduction}
Entanglement is a valuable property of some quantum states. In the
last few years attention was called to its time-evolution in the
context of quantum open systems \cite{history}. In special, it has
been considered a counterintuitive fact that for some systems and
initial states, entanglement can vanish in finite time, while
coherences only disappear asymptotically. Intuition was put in the
right perspective in the paper \cite{geo}, with a geometrical
picture of entanglement dynamics given in terms of the position of
asymptotic states with respect to the important subsets of entangled
and separable quantum states. Despite the classification problem be
completely characterized in this picture, the existence of examples
for some classes stayed unclear until the recent reference
\cite{examples}.

The present contribution will take the opportunity to state this
results in a more general form, to discuss the deepness and also to
clarify a new probabilistic approach to the question: {\it{given a
quantum system what will be the typical behavior of entanglement?}}
Finally, we make some considerations on the more ambitious
project, related to the so called Palis Conjecture, discussing what
would be the behavior of a typical open quantum system.

\section{Statement of the Result}
For the sake of clarity, we will state our main result as a
Theorem\footnote{Pretty much inspired on Prof. I. Volovich, who
would usually ask to a contributor: ``what is your theorem?''}.
Despite the fact that the majority of the discussions be made using
the two-qubit context, the results are much more general, as we may
discuss latter on in this paper.

Let us prepare the context first. We call $D$ the set of all
quantum states (in the sense of density operators) of a given
system. This closed convex set has a (closed and convex) subset $S$,
of separable quantum states, and the complement $E = D \setminus S$
is made of the entangled states. As we are concerned with time varying
quantum states (which can be considered as curves on $D$), it
becomes important to consider a trichotomy $\left\{int\, {S},
\partial S, E\right\}$, where $int\, S$ denotes the interior of the
set $S$, which sometimes we call deeply separable states\footnote{Here we restrict our attention to the finite dimensional case, where it is known\cite{BZ} that $S$ has positive measure when $D$ is given the normalized measure $1$.}, $\partial
S$ is the boundary of the set $S$ (relative to $D$) and we remember
that $E$ can be considered as an open set relative to $D$ (i.e. any
point of $E$ has a neighborhood of points of $D$ totally contained in
$E$).

When a quantum system evolves in contact with a reservoir, it is
typical to have a set (unitary\footnote{Unitary here, and throughout the paper, means with only one element.} or not) of asymptotic states. Let us
denote this set by $A$. The behavior of the time evolution of
entanglement can be classified and characterized in terms of the
relative position of such sets. The situation can vary depending on
the cardinality of $A$ (denoted $\left|A\right|$). The results are summarized in
the following classification and existence theorem:
\begin{quote}{\bf{Theorem}}
With the above definitions of $D$, $S$, $int\, S$, $\partial S$, $E$
and $A$, any quantum open system with linear dynamics and nontrivial set $A$ of
asymptotic states belongs to one, and only one, of the following
classes
\begin{enumerate}
\item $A = \left\{\rho_a\right\}$ with $\rho _a \in int\, S$;
\item $A = \left\{\rho_a\right\}$ with $\rho _a \in \partial S$;
\item $A = \left\{\rho_a\right\}$ with $\rho _a \in E$;
\item $\left|A\right| > 1$ with $A \subset int\, S$;
\item $\left|A\right| > 1$ with $A \cap \partial S \neq \emptyset$;
\item $\left|A\right| > 1$ with $A \subset E$.
\end{enumerate}
Moreover, examples of quantum systems can be obtained for each of these classes.
\end{quote}

\section{Idea of the proof}
 The proof of the classification part of the Theorem is simple. If $A$
is nontrivial, it must be unitary or larger. The first three classes
correspond to the unitary case, and each class corresponds to the
only element $\rho_a$ belonging to one of the three mutually
exclusive alternatives $int\, S$, $\partial S$, and $E$. For the
larger $A$ situation it is important to recognize that linearity of
quantum evolution implies a convex (hence connected) $A$. The first
question is whether $A \cap \partial S$ is empty. In negative case,
the system belongs to the fifth class, in affirmative case, in order
to avoid $\partial S$, $A$ must be contained in one of the
alternatives: $int\, S$ or $E$.

The existence part is constructive. We explicitly give examples for
each class, as can be checked on ref. \cite{examples}. It is
interesting to note that for the construction of the examples it was
sufficient to consider two qubits; however, it was necessary to allow
non-autonomous systems.

\section{Interpretation and Generality}
First of all, the geometric interpretation of entanglement time
evolution demystify the phenomenon of sudden death of
entanglement. If $A \subset int\, S$, any initial state needs
to lose entanglement in finite time in order to approach the
asymptotic set.

Moreover, the classification scheme also allows for the counterpart
of this phenomenon, whenever $A \subset E$. In such a case,
irrespective of the initial presence of entanglement, after a finite
amount of time there will be some entanglement.

The more interesting situations are the intermediate ones. The
situation $A =\left\{ \rho _a\right\}$, $\rho_a \in \partial S$ has
already received special attention in the literature \cite{Santos},
with the simple and interesting example of two independent qubits
spontaneous decaying. Depending on the initial state (even with the
same amount of initial entanglement) entanglement can die in finite
time or only asymptotically. This is vary natural, since $\rho_a$
can be approached both, by $\rho\left( t\right) \in S$ (``sudden
death'') or $\rho\left( t\right) \in E$ (``eternal life''). In such cases
($A \cap
\partial S \neq \emptyset$), an important refinement of this study
can be made case by case: if both situations are possible, which one
is prevalent? In what proportions do they happen? This can be well
defined \cite{examples} in terms of the measure of the sets of
initial states with each foreseen fate.

It must be clear that, despite the fact that much of our discussion
focus on two-qubit examples for simplicity, our results apply as
well for higher dimensionality and for the multipartite context,
when we can define more than one kind of entanglement. In this case,
$S$ must be considered as the states which do not have the specific
kind of entanglement one wants to consider and all the previous
reasoning apply. Of course, quantitative statements vary. Also, it is
interesting to recognize that for the multipartite scenario, one can
have a hierarchy of entanglements, and very interesting pictures can
be draw including cases when all entanglements vanish together
\cite{GHZ} or the more general situation when the family lose one
member per time.

\section{More speculative comments}

 An even more ambitious plan wouldbe to relate this study of time
evolution of entanglement to the so called {\emph{Palis
Conjecture}} on dynamical systems \cite{Palis}. In a precise way,
Palis enunciate that the vast majority of hyperbolic dynamical
systems has a finite number of attractors, whose basins gives a
set of total measure in the phase space.

Here, as quantum mechanical evolutions are simpler than general
hyperbolic dynamics on manifolds, we could dream of parameterizing
the set of possible evolutions for a well defined system (e.g. two
qubits), and of finding a meaningful measure on such a set, in order
to give quantitative answers to the question {\emph{what is the
typical behavior of a quantum system?}}


\begin{theacknowledgments}
MOTC wants to thank the organizers for the special meeting they
hold. Fapemig is acknowledged for founding the research and the
travel. RCD thanks financial support from CNPq.

\end{theacknowledgments}

\end{document}